\documentclass[twocolumn,secnumarabic,amsmath,amssymb,floatfix]{revtex4}
\usepackage{graphicx}% Include figure files
\usepackage{subfigure} %% fuer \subfigure
\usepackage{amssymb}

\begin{document}
%\preprint{APS/123-QED}

\title{Failure of feedback as a putative common mechanism of
  spreading depolarizations in migraine and stroke }
% Generation of spreading depolarizations by a failure of intrinsic feedback

\author {Markus A. Dahlem}
\author {Felix M. Schneider}
\author {Eckehard Sch\"oll}

\affiliation{Institut f{\"u}r Theoretische Physik, Technische Universit{\"a}t Berlin,  Hardenbergstra{\ss}e 36, D-10623 Berlin, Germany}

\begin{abstract}

The stability of cortical function depends critically on proper
regulation.  Under conditions of migraine and stroke a breakdown of
transmembrane chemical gradients can spread through cortical tissue.
A concomitant component of this emergent spatio-temporal pattern is a
depolarization of cells detected as slow voltage variations.  The
velocity of $\sim3$ mm/min indicates a contribution of diffusion.  We
propose a mechanism for spreading depolarizations (SD) that rests upon
a nonlocal or non-instantaneous feedback in a reaction-diffusion
system.  Depending upon the characteristic space and time scales of
the feedback, the propagation of cortical SD can be suppressed by
shifting the bifurcation line, which separates the parameter regime of
pulse propagation from the regime where a local disturbance dies out.
The optimisation of this feedback is elaborated for different control
schemes and ranges of control parameters.

\end{abstract}

\keywords{}
\pacs{}
\maketitle

{\bf During migraine and stroke neurological symptoms occur
  representing pathological events that spread through the cerebral
  cortex.  While these clinical observations have been known for a
  long time, only recently direct measurements were made. Two studies
  have revealed common spatio-temporal wave patterns, one using
  functional magnetic resonance imaging in a migraine patient
  \cite{HAD01} and another using electrodes placed directly on the
  exposed cortical surface to record electrical activity in a stroke
  patient \cite{FAB06}. The observed spatio-temporal patterns in the
  cortex constitute examples of excitable behavior that evidently
  emerges from pathological pathways. Spatial systems that exhibit the
  emergent property that activity breaks away from a local stimulation
  site are called {\em excitable media} \cite{Win91,MIK06}. The
  capacity to propagate pulses is the distinguishing feature of
  excitability in spatial systems.  As a mechanism for
  shifting the onset of excitability in a reaction-diffusion system we
  propose failure of nonlocal or non-instantaneous feedback control.  }

\section*{\sf I. INTRODUCTION}

The propagation of pathological states is a particular aspect
within the complex bidirectional relation between migraine and stroke
\cite{BOU05}.  Here we investigate this aspect, in particular, how
cortical tissue when modeled as an excitable medium becomes
susceptible to spreading events.  Our knowledge about the mechanisms
of propagation is still incomplete. It is generally believed that a
common reaction-diffusion process, called {\it cortical spreading
  depression} (CSD) \cite{LEA44,GRA63,GAR81,LAU94,Mar00a}, captures
essential features of the observed spreading phenomena during migraine
and stroke.  What makes cortical tissue susceptible to CSD has not
been determined. Since the smooth lissencephalic cortex of animals is
much more susceptible to CSD than the convoluted cortex of human, it
was suggested that CSD in humans occurs very close to the onset of
this emergent property \cite{DAH04a,DAH04b}.

\begin{figure}[!tb]
\centerline{\includegraphics[width=0.75\columnwidth]{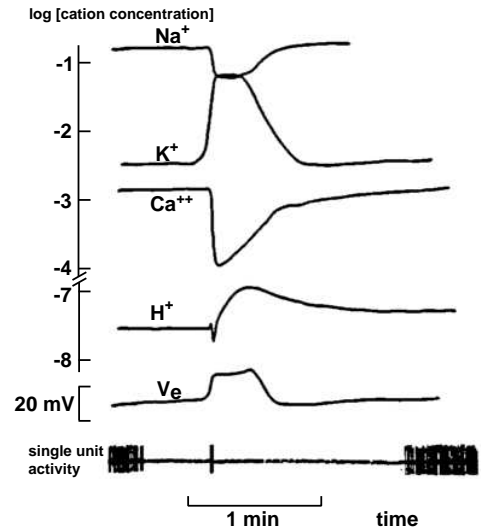}}
\caption
{\label{fig:sdcharakter} Classical measurement of electrophysiological
  changes during cortical spreading depression~\cite{LAU94}.  Upper
  four traces: logarithmic representation of changes in $Na^+$, $K^+$,
  $Ca^{++}$, and $H^+$ concentration, lower two traces: extracellular
  potential shift $V_e$, and recording from a single neuron (single
  unit activity) vs. time. (Modified  from \cite{LAU94})}
\end{figure}

CSD is locally characterized by a nearly complete temporary breakdown
of {\it ion homeostasis}, i.\,e., a stable stationary state
(Fig.\ \ref{fig:sdcharakter}) \cite{LAU94}.  Most electrophysiological
changes follow a similar temporal course and return to normal after
about one minute . A prominent signal is the slow negative potential
shift $V_e$ during CSD. It led to the term {\it spreading
  depolarization} (SD) to classify this and related phenomena as
depolarization waves in the cerebral cortex \cite{DRE06}.  Example of
related phenomena are SD waves that contribute to progressive
deterioration in regions surrounding an infarct core in stroke
patients. These SD waves are called periinfarct depolarizations (PID)
\cite{HOS96}.  We use the term SD for both cortical spreading
depression and periinfarct depolarizations in this paper. In this
terminology spreading depolarization is the more general term.  The
respective form of SD depends mainly on differences in the energy
state of cortical tissue. In particular, differences between CSD and
PID concern increased blood flow compensating the increased energy
demands, which is typical for CSD occurring in healthy tissue during
migraine but is reduced or missing in PID during stroke.

Differences in brain regions, e.\,g., concerning the cytoarchitecture
or the distribution of ion channel types, also modify SD.  We
disregard the detailed pathophysiological characterization of the
process, because it is still incompletely understood.  Instead, in
this study, SD is modeled with a standard reaction-diffusion system of
activator-inhibitor type.  Differences are reflected in the choice of
the parameter values of the system.  We extend the reaction-diffusion
model by different kinds of local feedback signals. We propose that
one way to think of such local feedback signals is in terms of control
\cite{SCH07}. In this view, the feedback  represents intrinsic cortical control
mechanisms that reduce cortical susceptibility for CSD by stabilizing
the physiological state of cortical tissue. Consequently, their
failure under certain pathological conditions leads to the emergence
of CSD, e.g., attributed to an underlying cortical hyperexcitability
in migraine \cite{WEL05} or due to low energy levels in ischaemic tissue during
stroke \cite{HOS96}.

In this study, different feedback mechanisms with characteristic time
and space scales are considered. Possible physiological basis of the
feedback signals are motivated and discussed. We find that the
excitability of the reaction-diffusion system can either increase or
decrease. For each considered feedback scheme mainly the sign of the
coupling strength determines whether feedback can suppress wave
propagation, i.e., stabilize the homogeneous steady state of the
tissue. A specific feedback scheme, in which activator and inhibitor
variables are cross-coupled by non-local connections, indicates that
opposed signs in the coupling strength for short-range and long-range
connections are favorable for controlling the homogeneous steady
state. This is a typical neuronal network connectivity pattern called
{\em Mexican-hat connectivity}.

%A failure of feedback that reduces excitability can explain the
%increased susceptibility of cortical tissue for CSD.

\section*{\sf II. THE MODEL}

\begin{figure}[!tb]
\centerline{\includegraphics[width=\columnwidth]{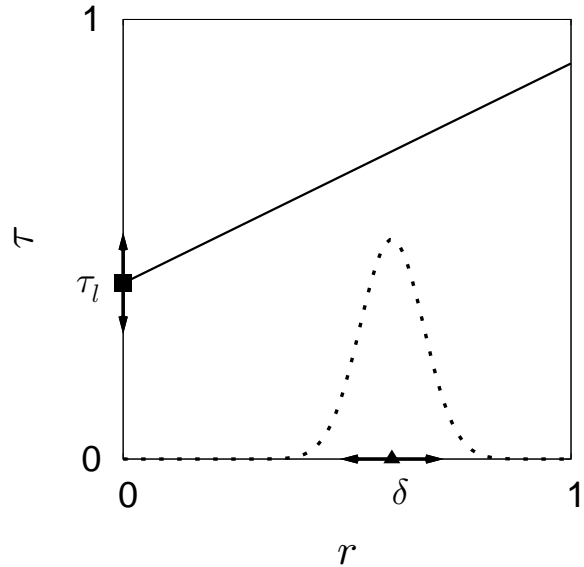}}
\caption
{Diagram of the spatial ($r$) and temporal ($\tau$) characteristics
  (solid line)
  of the feedback
  signal, which is used to extend the FitzHugh-Nagumo (FHN) reaction-diffusion
  model.  One of the two FHN variables is fed back via connections
  with length $\delta$. The length is defined by the radial coordinate
  $r$ of a local polar coordinate system, which lies parallel to the
  cortical surface and has its origin located where the connection
  terminates. Due to a finite propagation velocity $c_s$  and
  a latency time $\tau_{l}$ of the feedback signal, a total time delay
  $\tau=\tau_l+c_s^{-1}\delta$ occurs.  A kernel function $k(r)$ describes the
  spatial distribution of the connections, here exemplarily
  represented by a Gaussian (dashed).  Two physiologically
  motivated limit cases of this general feedback type are considered:
  $\blacktriangle$ instant feedback along long-range connections, and
  $\blacksquare$ local time-delayed feedback.
\label{fig:signal}}
\end{figure}

While all evidence suggests that a reaction-diffusion process captures
the essential features of SD, it is also clear that some features are
oversimplified if we model SD with a standard reaction-diffusion
mechanism of activator-inhibitor type.  To create a more accurate
albeit still generic model, we extend the standard reaction-diffusion
model by a feedback mechanism. We introduce the feedback signal
pathway in two variations.  First, we introduce long-range connections
as a spatial coupling in addition to diffusion. This is motivated by
the fact that high-frequency spikes in the population activity were
observed up to millimeters ahead of the approaching front of SD
\cite{HER94,LAR06}.  Second, we consider a localized time delayed
feedback in the cortex, which can account for a slow feedback signal
originating from the neurovascular coupling. Mathematically, these two
extensions can be viewed as two limit cases of a general signal
pathway (Fig. \ref{fig:signal}).  Such a pathway serves as an
additional coupling mechanism that is not included in a standard
reaction-diffusion model. With this approach, we aim to describe
universal features of excitable media modified by a feedback loop.

\subsection*{\sf A. Reaction-Diffusion Model}

We model the propagation of an initially localized breakdown of ion
homeostasis in cortical tissue (Fig.\ \ref{fig:sdcharakter})by an
activator $u$ and an inhibitor $v$ as dynamic variables. They are
coupled by kinetic reaction rates $f(u,v)$ and $g(u,v)$. We assume
that only the activator species diffuses in the medium. The equations
are
\begin{eqnarray}
\label{eq:udot}
 \frac{\partial u}{\partial  t}&=& f(u,v) + D
\nabla^2u  \\
\label{eq:vdot}
\frac{\partial v}{\partial  t}&=& \epsilon g(u,v).  
\end{eqnarray}
The parameter $\epsilon$ is the time scale ratio in the local dynamics
of activator and inhibitor variables.  The local spatial coupling is
introduced by the diffusion term in Eq.~(\ref{eq:udot}) with the
diffusion coefficient $D$.  Without loss of generality, the value of $D$
is normalized to unity.  The choice of this value only scales the spatial
coordinate.

In principle, there are two approaches to obtain the reaction rates
$f(u,v)$ and $g(u,v)$. In a bottom-up approach, the reaction rates
must be derived from a microscopic biophysical model of SD, where all
major electrophysiological properties are represented.  There is no
consent so far on the mechanism of SD.  Some models aim to provide a
complete but rather local description of SD \cite{TUC78,KAG00,SHA01},
that is, for processes occurring in a single neuron and its
surrounding compartments.  These models often introduce more than 20
dynamical variables.  Therefore, a bottom-up approach with a
subsequent reduction to two major activator and inhibitor agents seems
not to be amenable.  Nevertheless, it was shown that a two-species
activator-inhibitor model can reproduce the spatio-temporal pattern of
SD \cite{REG96,DAH04a,DAH04b}.  Even a reaction-diffusion model with a
single species was successfully introduced for the purpose of an
order-of-magnitude estimate of the expected propagation velocity of SD
\cite{GRA63,GAR81}. Such calculations follow essentially a top-down
approach, which we also adopt here.  The basic activator-inhibitor
mechanism, which we describe in the next paragraph, can be further
expanded (top-down) to facilitate detailed investigation of further
pathways and variables relevant to the study of specific questions
concerning SD. Successful examples of such an approach
are computational studies \cite{REV98,RUP99} that
 supported a controversial
hypothesis, namely that
cortical tissue surrounding an infarct core dies because of the
metabolic stress imposed by multiple SD
waves.

%\cite{MON06}

%We follow a similar stategy, We aim to describe generic
%spatio-temporal patterns and their control in an simple
%activator-inhibitor model.

The variables $u$ and $v$ assume the roles of activator
and inhibitor, respectively.  Their kinetic functions $f(u,v)$ and
$g(u,v)$ are given by a cubic nonlinearity and a linear function, respectively,
\begin{eqnarray}
\label{eq:f}
f(u,v)&=& a(u - \frac{u^3}{3}) - v  \\
\label{eq:g}
g(u,v)&=& u - \beta - \gamma v
\end{eqnarray}
with parameters $a$, $\beta$, and $\gamma$.  This is the
FitzHugh-Nagumo (FHN) system, which is widely used as a generic model of excitable media
\cite{Fit61,Nag62,LIN04}. Before we introduce the feedback, we compare
the simulated spatial-temporal patterns with the ones observed during
SD. In particular, the pulse profile and velocity in the FHN system are
related to the corresponding quantities in SD
(Fig.~\ref{fig:sdcharakter}). Following this approach, we can
estimate realistic values of the parameters of the FHN system.

\subsection*{\sf B. FitzHugh-Nagumo system with feedback}

To extend the standard FHN reaction-diffusion model we assume that
a nonlocal feedback signal $s(x,y,t)$  is coupled back into the medium
at any
point $(x,y)$ as
\begin{eqnarray}
\label{eq:s}
s(x,y,t)&=& K \int_{0}^{2\pi}\int_{0}^{\infty} \!\!  k(r)\,\, \\ \nonumber
&& \left(w(x+x_r,y+y_r,t-\tau)- w(x,y,t)\right)\,\, dr 
   \,\,d\phi
\end{eqnarray}
where $r$ and $\phi$ are the radius and the azimuthal angle,
respectively, of a local polar coordinate system in $(x,y)$ that lies
parallel to the cortical surface, i.\,e., $x_r=r\cos\phi$, and
$y_r=r\sin\phi$. The variable $w$ may be chosen as either the
activator $u$ or the inhibitor $v$.  The function $k(r)$ is the kernel
function describing the spatial distribution of the pathway, which is
typically peaked at a distance $r=\delta$, and $K$ is the coupling
strength.  The parameter $\tau=\tau_{p}(r)+\tau_{l}$, with
$\tau_{p}(r)=r/c_s$, is a time delay composed of a latency $\tau_{l}$
and a propagation delay $\tau_{p}$. The latter is due to the finite
signal propagation velocity $c_s$.  A schematic diagram of the signal
$s(x,y,t)$ is shown in Fig.~\ref{fig:signal}.  The feedback is, on the
one hand, characterized by the parameter $\tau$, and the kernel
function $k(r)$, i.\,e., $\delta$. We call this characterization the
{\it type of coupling } of the feedback signal.  On the other hand,
there are two choices of $w$, namely the activator $u$ and the
inhibitor $v$. Furthermore, each type can either be fed back to the
activator $u$ or inhibitor $v$ rate equation, i.\,e.,
Eq.~(\ref{eq:udot}) or Eq.~(\ref{eq:vdot}), respectively.  This allows
for four different combinations, named {\it schemes}, two
self-coupling and two cross-coupling schemes for each type of
coupling. The two self-coupling schemes are referred to as ``$uu$''
if $w=u$ and ``$vv$'' if $w=v$, and the two cross-coupling schemes
are referred to as ``$uv$'' if $w=u$ and ``$vu$'' if $w=v$.  For
example, in coupling scheme $uu$ the activator is used to compose the
feedback signal ($w=u$), which in turn is included in the activator
rate equation (i.e., self-coupling). In a vector short-hand notation, the FHN
system  
\begin{eqnarray}
\label{eq:short-hand1}
\frac{\partial \zeta}{\partial t}=F(\zeta),\,\,\,\,\mbox{with} \,\,\,\,\zeta=\left({u \atop {v}}\right)
\end{eqnarray}
is replaced by
\begin{eqnarray}
\label{eq:short-hand2}
\frac{\partial \zeta}{\partial t}=F(\zeta)+\sigma(x,y,t),
\end{eqnarray}
where the coupling is given by 
\begin{eqnarray}
\label{eq:short-hand3}
\sigma=\left({s \atop {0}}\right),\,\,\,\, \mbox{or} \,\,\,\,\sigma=\left({0 \atop {s}}\right)
\end{eqnarray}
with $w=u$ or $w=v$.
%, which defines the coupling matrix 
%\begin{eqnarray}
%\label{eq:short-hand4}
%\left(
%\begin{array}{*{2}{c}}
%uu & uv \\
%vu & vv \\
%\end{array}
%\right).
%\end{eqnarray}

Note that we have constructed the nonlocal feedback signal
(\ref{eq:s}) such that it contains the difference between the remote
and local values of $w$. As a consequence, the signal $s(x,y,t)$ tends
to zero if the pattern is homogeneous and stationary.  The motivation
for such a feedback is that this choice eliminates the feedback signal
in the case of successful suppression of wave propagation.  The signal
$s(x,y,t)$ can be viewed as an intrinsic noninvasive control because
it preserves the homogeneous steady state as the fixed point of the
uncontrolled ($K=0$) FHN system.  In this way, it can be compared with
the common time-delayed feedback control method introduced by Pyragas
\cite{PYR92}. This method has been widely used with great success in
problems in physics, chemistry, biology, and medicine \cite{SCH07}
including reaction-diffusion systems \cite{BEC02,BAB02,UNK03,SCH03a,SCH06c}.
In particular, it was demonstrated that it can be used to control the
coherence and the timescales of noise-induced oscillations in a single
FHN system \cite{JAN04,BAL04,PRA07} and in two coupled excitable FHN
systems \cite{HAU06,HOE07} as well as noise-induced patterns in
reaction-diffusion systems \cite{STE05,STE05a,HIZ06,HIZ08} and wave
propagation in excitable media \cite{BAL06}. This motivates our efforts to investigate
whether a failure of such an intrinsic noninvasive control scheme can
explain the onset of excitation spread in a spatially continuous FHN
systems as a model of spreading pathological processes in the cortex
during migraine and stroke.

%PYR93,BIE94,PIE96,HAL97,SUK97,LUE01,PAR99,KRO00a,FUK02,LOE04,ROS04a,POP05,SCH06a
%Next we consider two limit cases of this additional coupling scheme.

\subsubsection*{\sf 1. Instant feedback along long-range connections}

\begin{figure}[!tb]
\centerline{\includegraphics[width=1.0\columnwidth]{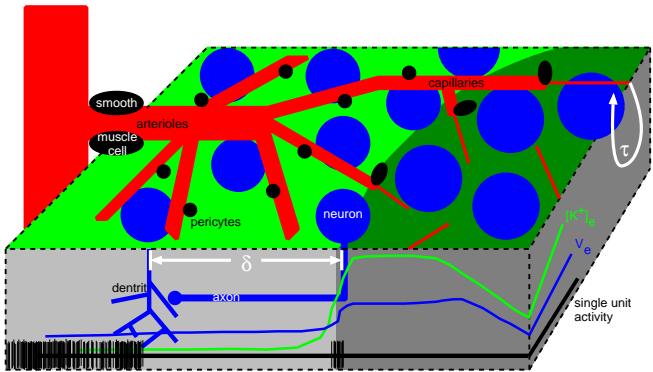}}
\caption
{Scheme of the cortical surface during cortical spreading depression.
  Red: blood vessels formed by endothelial cells. The cerebral blood
  flow starts at arterioles and leads to smaller capillaries, forming
  the capillary bed.  Blue: neurons, green: glial cells. A pulse of
  cortical spreading depression moves from right to left, indicated by
  the shadowed region. The extracellular concentration changes of
  potassium $[K^+]_e$, the extracellular potential shift $V_e$, and
  the single unit activity are schematically related to the pulse on
  the lateral front surface. Note that the size of the different cells
  types are enlarged by about a factor of 10 in relation to these
  measured signals. The two considered feedback mechansim are also
  illustrated: (i) long-range instantaneous lateral connections of
  length $\delta$ between neurons (shown on the lateral front
  surface), (ii) a time delayed feedback control ($\tau$) of Pyragas
  type due to local response of the neurovascluar unit (indicated on
  the right lateral side surface).
\label{fig:nvc}}
\end{figure}

The first type of coupling that we consider is a limit case of
Eq.~(\ref{eq:s}). It is motivated by the observation of increased
neuronal activity millimeters ahead of the approaching SD
front. Although the cause and the effect of this activity remains
unclear, it can be assumed that such activity is induced by SD via
lateral cortical connectivity patterns (Fig.~\ref{fig:nvc}). In this
case the signal propagation velocity $c_s$ is several orders of
magnitude faster than the velocity $c$ of SD, i.\,e., \mbox{$c_s\gg
  c$} ($50 \mu\mbox{m s}^{-1}$ in physical units). Consequently, the
propagation delay $\tau_{p}$ can be neglected.  Furthermore, the time
scale of typical changes in electrophysiological activity is much
faster than the time scale of changes due to the breakdown of ion
homeostasis. The latter time scale is of the order of seconds
(Fig.\ \ref{fig:sdcharakter}), which is also in agreement with our
simulations of the FHN system if we dimensionalize the model equations
(see section III below), while the former time scale, i.\,e., the
delay due to chemical or electrical transmission at synapses, is of
the order of milliseconds. Therefore, it is reasonable to assume that
the latency $\tau_l$ of $s(x,y,t)$ is much smaller than the time scale
of SD and that $\tau_l$ can also be neglected.  This assumption can
fail for metabotropic ion channels, like metabotropic glutamate
receptors, which have increased open probabilities in the range of
seconds after their activation.  However, for ionotropically mediated
activity we assume that $\tau_p \approx 0$ and $\tau_l \approx 0$, and
therefore that the additional coupling signal $s(x,y,t)$ is, in this
case, an instantaneous process ($\tau=0$).  Furthermore, we assume
that the connectivity pattern is rather localized around the typical
coupling length $\delta$, and therefore, restricting ourselves to one
spatial dimension, $x$, we approximate the kernel function by
$\delta$-functions $k(r) \approx \delta(x-\delta)+\delta(x+\delta)$.
Taking these limits, Eq.~(\ref{eq:s}) reduces to
\begin{eqnarray}
\label{eq:ddc}
s_{\delta}(x,t) =   K (w(x-\delta,t) - 2w(x,t) + w(x+\delta,t))
\end{eqnarray}
Note that Eq.~(\ref{eq:ddc})  only in the limit $\delta \rightarrow 0$
becomes a standard diffusion term, however, for large $\delta$, as
considered here, it adds a novel type of nonlocal spatial coupling to
the FHN model.

\subsubsection*{\sf 2. Local time-delayed feedback}

% and associated tissue matrix proteins

The second type is also a limit case of Eq.~(\ref{eq:s}). 
We consider a scenario
that takes into account the coupling within the neurovascular unit.
Therefore, it is  particularly important for the dynamics of
periinfarct depolarizations during stroke progression.  The cerebral
blood flow (CBF) is tightly regulated by the neurovascular unit to
meet the brain's metabolic demands. These demands are extremely high
during SD.  The dynamics of the neurovascular unit are governed by
various different cells types (Fig.~\ref{fig:nvc}). The interaction takes place between
endothelial cells building the vessel walls, mural cells controlling
vessel diameter, and glia and neurons.  For the purpose of our
simplified scheme, that is, describing universal features of
reaction-diffusion models of SD modified by an additional signal
pathway, we concentrate on some key features of blood flow regulation
that occur within the time and space scale given by a single passage
of SD and that mimic the influence of the neurovascular unit.

The local CBF starts at arterioles, that is, blood vessels that
extend and branch out from an artery and lead to capillaries.
Capillaries are the smallest vessels, measuring $5-10 \mu\mbox{m}$ in
diameter. They form the capillary bed, a local network supplying the
brain tissue.  The CBF is regulated by two types of mural cells.
Smooth muscle cells regulate arterioles, and pericytes regulate
capillaries.  The majority of the innervation of cerebral blood
vessels terminate near capillaries suggesting that blood flow control
is initiated in capillaries \cite{PEP06}.  Therefore, we need to
consider that increased activity during SD produces an initially
localized hemodynamic response evoked by pericytes.  To describe the
action of the neurovascular unit on SD, the temporal and spatial
distribution of the hemodynamic response needs to be considered.

First, we consider the spatial distribution of the hemodynamic
response, which is determined by the vascular architecture.  A precise
spatial coordination of segmental vascular resistance is needed to
effectively increase blood flow in a larger cortical area.  Propagated
vascular response signals are utilized to achieved this. It was shown
that after the pericytes evoke a local capillary constriction, a pulse
of constriction propagates at about $2 \mu\mbox{m\,s}^{-1}$ to distant
pericytes \cite{PEP06}. Furthermore, a dilatory signal mediated by
release of vasoactive agents propagates from metabolically active
cells and evokes a remote response in upstream precapillary arterioles
\cite{IAD97}.  Considering the slow velocity of $2 \mu\mbox{m\,s}^{-1}$
we can assume that global effects mediated by arterioles take place
behind SD, i.\,e., after its passage, and therefore do not have a
direct influence on the propagation of the wave front having passed.
We assume that the response of the capillaries is rather localized, so
that we can approximate the kernel function $k(x)\approx \delta(x)$.
Consequently there is only a local response. Still
there exists a latency delay $\tau_l$ due to slow metabolic effects
when a local hemodynamic response is evoked by pericytes.  Taking
these limits, Eq.~(\ref{eq:s}) now reduces to
\begin{eqnarray}
\label{eq:tdc}
s_{\tau}(x,t) =   K (w(x,t-\tau) - w(x,t)),
\end{eqnarray}
which is identical with the noninvasive time-delayed feedback signal
first introduced by Pyragas \cite{PYR92} for chaos control. This
control method will now be applied to travelling pulses in excitable
media.  The latency delay time $\tau$ will be varied in our
simulations.

Note that during stroke progression, the recruitment of periinfarct
tissue into the infarct core is probably mediated by wave trains of
periinfarct depolarizations \cite{DRE06}. Such periodic wave patterns
are likely to be influenced by large scale hemodynamic responses.
After the leading SD has initiated a slow propagating wave of
constriction \cite{PEP06}, subsequent waves in the wave train
formation can be influenced by this.  This is beyond the scope of our
study.  To model a single SD pulse, we include the neurovascular unit by
a localized time-delayed feedback signal $s_{\tau}(x,t)$ evoked from
the change in blood flow which in turn is caused by pericytes in
response to changes in neural activity.

% precapillary 

\section*{\sf III. RESULTS}

%Excitable media exhibit the emergent property that activity breaks away
%from a local stimulation site \cite{Win91,MIK06}. This is the
%distinguishing feature of excitability in spatial systems.  

% match the pulse profile and its
%velocity in the FHN system with the corresponding quantities observed
%during SD . This allows us to 

It was suggested that SD in humans occurs close to the bifurcation of
the onset of spreading activity in excitable media
\cite{DAH04a,DAH04b}.  Therefore, we first determine the location of
this onset in the FHN system.  Then we estimate the parameters of the
FHN system from experimental data on SD (Fig.\ \ref{fig:sdcharakter})
and compare them with corresponding quantities in brain
tissue. Finally, we investigate how the onset of excitability is
influenced by the two types of coupling, $s_{\delta}(x,t)$ and
$s_{\tau}(x,t)$, introduced in the previous section, in particular,
how the onset is influenced by the newly introduced spatial and
temporal scales $\delta$ and $\tau$, respectively. To achieve this, we
first consider the uncontrolled FHN system in the regime where
sustained pulse propagation is possible and investigate whether or not
the additional coupling schemes suppress pulse propagation. In this
context, we can view the coupling scheme as a control scheme with the
control goal to stabilize the homogeneous steady state
\cite{HOE05,SCH06a,DAH07}. Second, we choose parameter pairs
$(K_0,\delta_0)$ and $(K_0,\tau_0)$ that suppress pulse propagation and
determine the location of the onset of propagation in the FHN system
with these coupling schemes. Furthermore, we determine the location of
the onset of propagation in the FHN system for those cases where the
coupling schemes cannot suppress pulse propagation.

%Second, we set the FHN system in the regime
%where sustained wave propagation is not supported, i.\,e., the FHN
%system is unexciatble, and investigate whether the additional coupling
%schemes can let to the emergence of sustained wave propagation.

\subsection*{\sf 1. Boundary of pulse propagation}

\begin{figure}[!tb]
  \centerline{\includegraphics[width=\columnwidth]{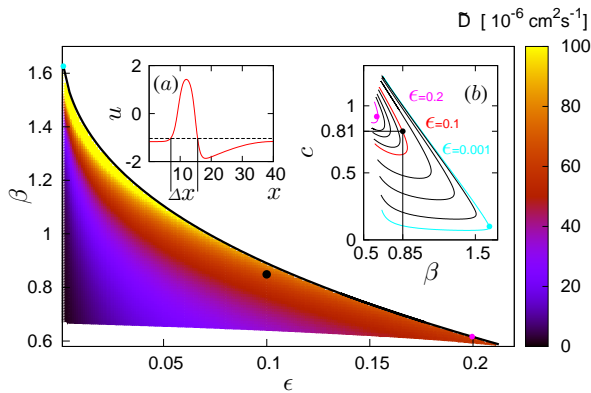}}
\caption
{ Parameter space of the FHN system at $\gamma=0.5$ and $a=1$.  The
  dimensionalized diffusion coefficient is color-coded as a function of
  the time scale ratio $\epsilon$ and the
  excitability threshold parameter $\beta$. The
  excitable regime (color-shaded) and the non-excitable regime (white)
  are separated by the boundary of propagation (thick black
  line). Above this boundary traveling pulse solutions do not
  exist. Inset (a) shows a typical pulse profile corresponding to the
  black dot in the main figure. Inset (b) shows homoclinic orbits of the
  FHN system in the co-moving frame with velocity $c$ as a function of
  $\beta$ for different values of $\epsilon$. In this diagram, the
  location of the boundary of propagation is determined by the value
  of $\beta$ at the turning point of each curve and the corresponding
  $\epsilon$ value, exemplarily shown by cyan and magenta dots for
  $\epsilon=0.001$ and $\epsilon=0.2$, respectively, in inset (b) and
  the main figure. }
\label{fig:pb}
\end{figure}

The spread of an initially local activity arises in the system
described by Eqs.~(\ref{eq:udot})-(\ref{eq:g}) if a critical parameter
value is crossed above which the medium is susceptible for sustained
propagating excitation patterns \citep{Mik91,Win91,Hak99,MIH02}.  In a
1D medium this border is obtained by finding  the
regime where stationary solutions exist in a co-moving frame for specific activator and
inhibitor profiles.  In the co-moving frame $\xi=x+ct$, the coupled partial
differential equations for activator and inhibitor variables
Eq.~(\ref{eq:udot})-(\ref{eq:vdot}) transform to a system of second
order ordinary differential equations for $U(\xi)$ and $V(\xi)$
\begin{eqnarray}
\label{eq:cmf}
u(x,t)&=&U(\xi)\nonumber\\
v(x,t)&=&V(\xi)
\end{eqnarray}
which can be further transformed to a system of coupled first order ordinary
differential equations of three variables, $U(\xi)$, $V(\xi)$,
$W(\xi)\equiv\partial U(\xi)/\partial \xi$ \cite{KUZ95}.  Due to
this transformation, $c$, the propagation velocity, is introduced as an
additional parameter.  In this system, homoclinic orbits can be found
using a continuation method \cite{DOE06}. They correspond to
travelling pulse solutions in Eq.~(\ref{eq:udot})-(\ref{eq:vdot})
(Fig.~\ref{fig:pb}, inset b).  The transition into a region where
homoclinic orbits exist marks a bifurcation of codimension one.  In
the parameter space of excitable media 
this
bifurcation separates
the
regime supporting travelling pulses from the non-excitable regime.  This
border  is shown in
Fig.~\ref{fig:pb} in the parameter plane $(\epsilon,\beta)$ at the
section $a=1$ and $\gamma=0.5$. As expected, the transition into the
non-excitable regime is achieved by increasing either $\beta$ or
$\epsilon$.  Increasing $\beta$ increases the threshold of the excitable
medium, while increasing $\epsilon$ changes the time scale ratio
between the dynamics of activator and inhibitor, with larger values of
$\epsilon$ corresponding to faster inhibitor dynamics.

%###########################      FS   START ###################################

\subsection*{\sf 2. Estimate of space and time scales}

% , we introduce spatial and temporal units, $[x]$ and $[t]$,
%respectively. This is done in two steps, the first

The main reason to introduce spatial and temporal units in our FHN
system is to obtain a better understanding of the typical values of
the spatial and temporal coupling scales, $\delta$ and $\tau$, respectively.
Furthermore, the value of the diffusion coefficient $D$ can be determined 
if we introduce dimensional units in the FHN system. 
Note that $D$ is not a bifurcation parameter. In the
non-dimensionalized FHN system Eqs.~(\ref{eq:udot})-(\ref{eq:g}), $D$ 
can assume any value without changing the dynamics in a qualitative 
way. For example, changing $D$ has no effect on the location of the
propagation boundary shown in Fig.~\ref{fig:pb}. Varying $D$ is equivalent
to scaling space units by $D^{-\frac{1}{2}}$.  Therefore, the physical
value of the diffusion coefficient $D$ can be determined if we match 
the patterns obtained in the FHN system with the ones observed experimentally 
during SD.

First, we define characteristic space and time units, $x_0$ and
$t_0$, respectively. This is done with respect to the spatio-temporal
patterns of the uncontrolled ($K=0$) FHN system. These characteristic
units are chosen such that the patterns calculated from the non-dimensionalized
Eqs.~(\ref{eq:udot})-(\ref{eq:g}) match the measured ones of Fig.~\ref{fig:sdcharakter}.
Introducing dimensional space (X) and time (T) variables by $x=X/x_0$ and $t=T/t_0$,
the FHN system Eqs.~(\ref{eq:udot}),(\ref{eq:vdot}) assumes its dimensional form 
with the dimensionalized diffusion coefficient $\tilde{D}\equiv x_0^{2}/t_0$.
%In the co-moving frame $x_0$ and $t_0$ are related by $x_0= c_0 t_0$ with the dimensional
%propagation velocity $c_0$. Hence $\tilde{D}= c_{0}^{2} t_0$

Now we compare the simulated pulse width $\Delta x$ and duration
$\Delta t$ with the measured pulse width $\Delta X$ and measured
duration $\Delta T$ (Fig.~\ref{fig:sdcharakter}, $\Delta T \approx
20s$).  Using the typical measured velocity of SD $C \approx 50 \mu m
s^{-1}$, we obtain $ \Delta X=C \Delta T \approx 0.1cm$.  The
simulated pulse width $\Delta x$ and duration $\Delta t$ are related
by $\Delta x= c \Delta t$ with the non-dimensionalized propagation
velocity $c$, which appears as a parameter in Eq.~(\ref{eq:cmf})
(e.g. for $\epsilon = 0.1$ and $\beta = 0.85$ : $\Delta x \approx 8.7$
in Fig.~\ref{fig:pb} (inset~a) and $c \approx 0.81$ in Fig.~\ref{fig:pb}
(inset~b), hence $\Delta t \approx 10.70$).  Hence we infer the space
and time units $x_0=\Delta X / \Delta x \approx 115 \mu m$ and
$t_0=\Delta T / \Delta t \approx 1.9s$, respectively, which yields a
diffusion coefficient $\tilde{D}=x_0^2/t_0\approx 70\cdot10^{-6} cm^{2}s^{-1}$,
which is a reasonable value.  Note that the pulse width $\Delta x$,
the velocity $c$ and hence $x_0$, $t_0$ and $\tilde{D}$ depend upon
the parameters $\beta$ and $\epsilon$. % (Fig.~\ref{fig:pb}).

In Fig.~\ref{fig:pb} the obtained values for the diffusion
coefficient $\tilde{D}$ for each pair ($\epsilon$,$\beta$) are shown
in color coding.  
It ranges from $5\cdot 10^{-6}-100\cdot
10^{-6}\mbox{cm}^{2}\mbox{s}^{-1}$ in the main part of the parameter
space.
%The obtained values for the diffusion
%coefficient $\tilde{D}$ range from $5\cdot 10^{-6}-100\cdot
%10^{-6}\mbox{cm}^{2}\mbox{s}^{-1}$ in the main part of the parameter
%space (Fig.\ \ref{fig:pb}).
As expected, low values are obtained in regions far away from the
boundary of propagation.  There the excitability of the system is
high. On the other side, close to the propagation boundary, the values
of $\tilde{D}$ are higher. In this regime, the medium is weakly
excitable or \emph{subexcitable}. The correlation between the value of
the obtained diffusion coefficients $\tilde{D}$ and the excitability
regimes requires some further clarification, which will be addressed
in the discussion. However, note that the values of $\tilde{D}$ in the
parameter plane $(\epsilon,\beta)$ at the section $a=1$ and
$\gamma=0.5$ are in the range of expected values for a diffusing
substance participating in the mechanism of SD, as discussed in the
literature \cite{GRA63,GAR81,NIC81,WIL99,DAH04b}.

%###########################      FS   END ###################################

\subsection*{\sf 3. Control of spreading depressions by $s_{\delta}(x,t)$ and $s_{\tau}(x,t)$}

How does the additional coupling  $s(x,t)$ change the FHN
system? To answer this, we perform simulations with a variety of FHN
systems in the parameter plane $(\epsilon,\beta)$ at $a=1$
and $\gamma=0.5$. The systems are chosen close to the propagation
boundary (Fig.~\ref{fig:pb}). We include both types of coupling, the
instant feedback along long-range connections $s_{\delta}(x,t)$,
described in section II.1, and the local time-delayed feedback
$s_{\tau}(x,t)$, described in section II.2.  Both types are
considered separately. For each type of coupling there are four coupling
schemes, two self-coupling schemes $(uu,vv)$ and two cross-coupling
schemes $(uv,uv)$.

To obtain the influence of the coupling schemes on the FHN system, we
start each individual simulation with a stable pulse profile of the
reaction-diffusion system obtained without the coupling signal
($K=0$). Then, for $K\neq 0$, we determined whether or not the pulse
propagation is terminated.  If the propagation is suppressed, we also
determine how long the pulse can still propagate before it disappears.
This distance defines the volume of tissue at risk (TAR), referring to
the risk of cortical tissue surrounding a local pathological core of
being recruited into the disturbed state \cite{DAH07a}.  This TAR
value is taken as a measure of the efficiency of the coupling scheme
as a control method.

\begin{figure}[!tb]
\centerline{\includegraphics[width=\columnwidth]{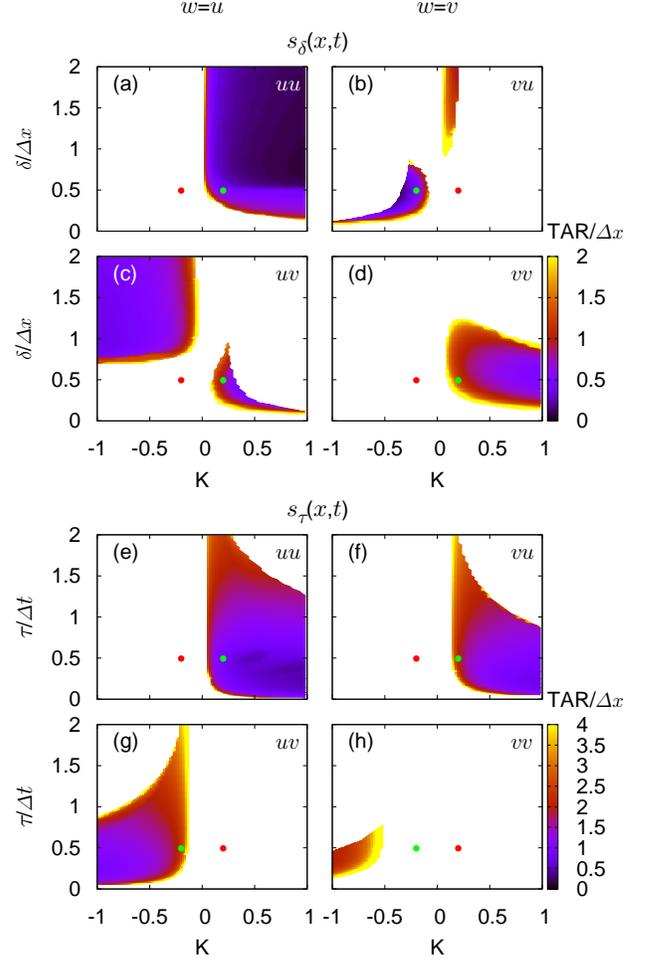}}
\caption
{Control planes for the
four coupling schemes of the two types of feedback $s_\delta(x,t)$
({\sf a-d}) and $s_\tau(x,t)$ ({\sf e-h}).  ({\sf a,e}):
self-coupling of the activator signal ($uu$); ({\sf b,f}):
self-coupling of the inhibitor signal ($vv$); ({\sf c,g}):
cross-coupling with the feedback signal composed of the activator
($w=u$) and fed back to the inhibitor rate equation ($uv$); ({\sf
  d,h}): cross-coupling with the inverse configuration as compared to
({\sf c,g}), i.\,e., ($vu$).  The parameter values of the FHN system
are $\epsilon=0.1$, $\beta=0.85$, $a=1$, and $\gamma=0.5$. This system
is close to the propagation boundary (see black dot in
Fig.~\ref{fig:pb}).  Regions in which pulse propagation is suppressed
by the feedback signal are shown color coded by the tissue at risk
(TAR) value, that is, the volume of tissue recruited into the
pathological state before the pulse dies out. Low TAR values indicate
that the feedback signal is more efficient.  Optimal values, i.\,e., low TAR values, for
$\delta$ and $\tau$ are mainly located at about $0.5$ times the activator pulse
width and duration, respectively (see text). 
Red and green dots mark control parameter values $(K,\delta)$ and $(K,\tau)$ for which the shift in the
pulse propagation boundary is calculated in Fig.~\ref{fig:pbs}.
\label{fig:kDeltaTau}}
\end{figure}

First, we consider the four non-local schemes in $s_{\delta}(x,t)$, i.\,e.,
instant coupling along long-range connections (Sec. II.1).  Regions in
which pulse propagation is suppressed are shown in
Fig.~\ref{fig:kDeltaTau} {\sf a-d}. The TAR value is given by a color
code in this regions.  For the two cases where we introduce self-coupling 
(Fig.~\ref{fig:kDeltaTau} {\sf a} ($w=u$) and {\sf d}
($w=v$)), mainly the sign of the coupling constant $K$ determines the
success of the coupling scheme. Pulse propagation cannot be suppressed
with negative values of $K$. This can be intuitively understood, if we
consider only the effect of the signal $s_{\delta}(x,t)$ on the
homogeneous steady state. A small disturbance from the homogeneous state is
destabilized by $s_{\delta}(x,t)$ if $K<0$ and stabilized if
$K>0$.  Optimal values of $\delta$ are obtained for $w=u$ at
$\delta\approx \Delta x$ and for $w=v$ at $\delta\approx 0.5 \Delta x$.  The
picture changes for the two schemes with cross-coupling
(Fig.~\ref{fig:kDeltaTau} {\sf b} ($w=v$) and {\sf c}
($w=u$)). Successful suppression of pulse propagation in these cases
depends on both the coupling strength $K$ and the coupling length
$\delta$. The change in the sign of $K$ occurs at $\delta\approx \Delta x$.
If $\delta < \Delta x$, $K$ must be positive (negative) if the coupling
term $s_{\delta}(x,t)$ is fed into the inhibitor (activator) balance
equation Eq.~(\ref{eq:vdot}) or (\ref{eq:udot}), respectively, and vice versa 
for $\delta \Delta x$. In other words, the short and
long range control domains have opposite signs of $K$.  Optimal values
of $\delta$ in the short range control domain are obtained at
$\delta\approx 0.5 \Delta x$.

Next, we consider the local time-delayed feedback coupling
(Sec. II.2). In each of the four coupling schemes control is only
achieved if $K$ is either positive or negative
(Fig.~\ref{fig:kDeltaTau} {\sf e-h}).  For positive $K$, the pulse
propagation is suppressed if the signal is coupled into the activator $u$. 
%by $\sigma=\left({s \atop {0}}\right)$, cf. Eqs. (\ref{eq:short-hand2})-(\ref{eq:short-hand3}).  
The two domains in the $(K,\tau)$ plane, where pulse propagation is
suppressed, are of comparable size (Fig.~\ref{fig:kDeltaTau} {\sf e}
(self-coupling) and {\sf f} (cross-coupling)).  For negative $K$, the
pulse propagation is suppressed if the signal is coupled into the inhibitor $v$.
%by $\sigma=\left({0 \atop {s}}\right)$. 
 In this case, the cross-coupling
scheme (Fig.~\ref{fig:kDeltaTau} {\sf g}) has a domain of comparable
size and shape as for cases shown in Fig.~\ref{fig:kDeltaTau} {\sf e,f}. The domain for the self-coupling scheme
is much smaller with only large values of TAR, i.\,e., less efficient
suppression of pulse propagation (Fig.~\ref{fig:kDeltaTau} {\sf h}).
For all four cases, the optimal delay time is $\tau\approx 0.5 \Delta
t$.

\subsection*{\sf 4. Shift in the onset of excitability  by $s_{\delta}(x,t)$ and $s_{\tau}(x,t)$}

We now investigate the shift in the location of the propagation
boundary (Fig.~\ref{fig:pb}) caused by the four coupling schemes of
 both   types of coupling $s_{\delta}(x,t)$ and
$s_{\tau}(x,t)$. This boundary separates the  excitable regime from the
non-excitable regime. Intuitively it is clear that the suppression of
pulses shifts the propagation boundary towards the excitable regime of
the system with $K=0$.  If we consider the coupling schemes as control
mechanisms, such a shift manifests itself as a successful control strategy that
reduces the excitability of the system.

\begin{figure}[!tb]
\centerline{\includegraphics[width=\columnwidth]{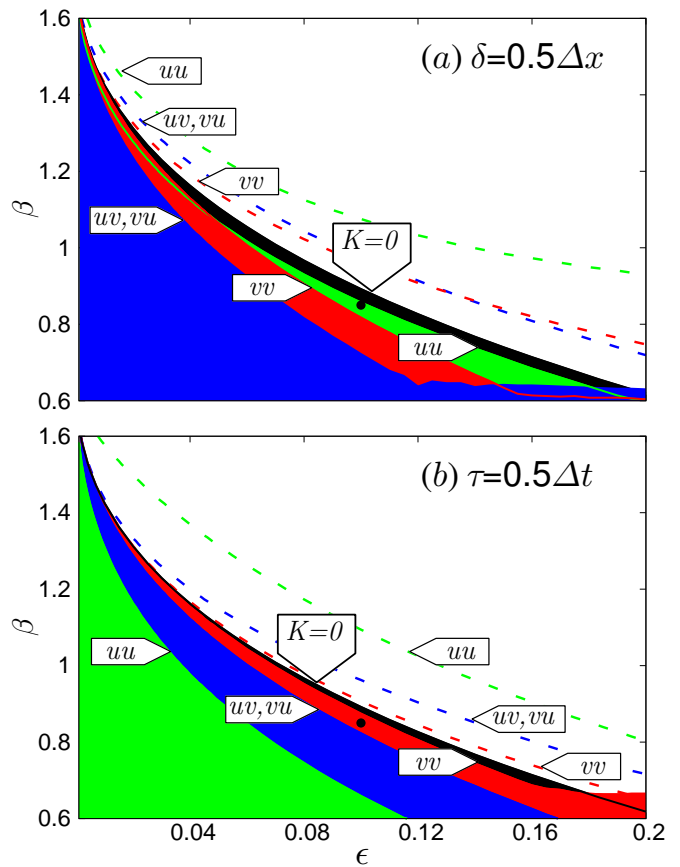}}
\caption
{ Parameter space ($\epsilon,\beta$) as in Fig.~\ref{fig:pb} showing excitable 
  and non-excitable regimes for (a) non-local, (b) local time-delayed coupling. For the uncontrolled system ($K=0$) the
  border between the excitable regime and the non-excitable regimes is
  between the black and white region. The other colored
  regions (green, red, blue) visualize the reduction of the excitable
  regime by the different coupling schemes 
  as indicated by the location of the green dots in Fig.~\ref{fig:kDeltaTau}), i.e.
  $K$ is chosen as $K=0.2$ (uu, uv, vv in (a); uu, vu in (b)), 
  or $K=-0.2$ (vu in (a); uv, vv in (b)). For each
  coupling scheme with opposite sign of $K$ (red dots in
  Fig.~\ref{fig:kDeltaTau}) the excitable regime
  increases as indicated by the dashed lines.}
\label{fig:pbs}
\end{figure}

First, we consider the non-local type of coupling $s_{\delta}(x,t)$.
The parameter values for $s_{\delta}(x,t)$ are set to $K=0.2$ and
$K=-0.2$, and $\delta=0.5\Delta x$.  For each of the four coupling
schemes with $\delta=0.5\Delta x$ only one value of $K$, either
$K=0.2$ or $K=-0.2$, suppresses pulse propagation (marked by green
dots in Fig.~\ref{fig:kDeltaTau}).  With this choice and depending
upon the four coupling schemes, the control signal $s_{\delta}(x,t)$
lies either close to optimal $\delta$ values or on the opposite side,
mirrored along the $K=0$ axis (marked by red dots in
Fig.~\ref{fig:kDeltaTau}), a location where the control goal is not
achieved.  In the cases of the two cross-coupling schemes, the value
$K=-0.2$ for $w=v$ (Fig.~\ref{fig:kDeltaTau} {\sf b}) and the value
$K=0.2$ for $w=u$ (Fig.~\ref{fig:kDeltaTau} {\sf c}) lies in the
center of the control domain with the short range connections.  If we
consider the four cases in which pulses are suppressed, the location
of the propagation boundary is shifted into the excitable regime,
indicated by differently colored regions in Fig.~\ref{fig:pbs} {\sf
  a}). It is evident that the $uv$ and $vu$ cross coupling schemes are
most efficient. Note that for the least efficient coupling scheme
  ($uu$) the propagation boundary (edge of the green region in
  Fig.~\ref{fig:pbs} {\sf a}) is actually not shifted beyond the point
  $\epsilon=0.1$ and $\beta=0.85$ (black dot). This seems to be
  contradictory with the control domain shown in
  Fig.~\ref{fig:kDeltaTau} {\sf a}) but is consistent with the definition
  of the propagation boundary as a bifurcation line. Beyond this
  bifurcation line there do not exist any traveling pulse solutions, whereas
  Fig.~\ref{fig:kDeltaTau} {\sf a}) only states that a specific
  traveling pulse solution with a specific amplitude 
  (the one of the uncontrolled system) is
  suppressed. This example illustrates the different nature of the
  information shown in Fig.~\ref{fig:kDeltaTau} and
  Fig.~\ref{fig:pbs}.   In the four cases with the opposite sign of
the parameter value $K$, i.\,e., where pulses are not suppressed, the
propagation boundary is shifted towards the non-excitable regime of
the uncontrolled ($K=0$) FHN system (dashed lines in
Fig.~\ref{fig:pbs} {\sf a}).

Second, we consider the local, time-delayed type of coupling $s_{\tau}(x,t)$.  
The parameter
values for $s_{\tau}(x,t)$ are set to $K=0.2$ and $K=-0.2$, and
$\tau=0.5\Delta t$.  To provide a comparison with the former type of coupling 
$s_{\delta}(x,t)$, we also choose the short range domain (as for $s_{\delta}(x,t)$  
in Fig.~\ref{fig:kDeltaTau} {\sf b,c})
as the reference point and use these  values of $K$ also for
$s_{\tau}(x,t)$.  As before they are marked by green dots in
Fig.~\ref{fig:kDeltaTau} {\sf e,f,g} if control is successful and
red otherwise. Only for the scheme 
%($uv$) 
($vv$) (Fig.~\ref{fig:kDeltaTau}
{\sf h}) both  types of coupling for $K=-0.2$ and $K=0.2$ lie outside the
control domain. Since the type with $K=-0.2$ is closer to a control
domain, this is marked green, and correspondingly its propagation
boundary is slightly pulled back as indicated by the red region in 
Fig.~\ref{fig:pbs} {\sf b}. Again we find a similar pattern as for the non-local 
type of coupling
$s_{\delta}(x,t)$. The coupling schemes that suppress pulse propagation
reduce excitability (colored regions in Fig.~\ref{fig:pbs} {\sf b}) and
the other ones increase excitability (dashed lines).  Note that also for the coupling
scheme ($vv$)
%($uv$) 
this pattern occurs, with the only difference that the
propagation boundary (edge of the red region in Fig.~\ref{fig:pbs} {\sf b})
is not shifted beyond the point $\epsilon=0.1$ and $\beta=0.85$ (black dot),
which is in agreement with the control domain shown in
Fig.~\ref{fig:kDeltaTau} {\sf g}.

%{\sf [Maybe here one further paragraph mentioning that in each
%    subfigure of Fig.~\ref{fig:pbs} two pairs of lines coincide (WE
%    DO NOT KNOW YET WHY?)  and that for large $\epsilon$ $(>0.1)$ the
%    propagation boundary may exhibit a sharp upwards bend, because it
%    collides with the oscillatory regime.]  }

\section*{\sf DISCUSSION AND CONCLUSIONS}

We have shown that a weakly excitable reaction-diffusion system can be
shifted across the boundary of pulse propagation (Fig.~\ref{fig:pbs})
by  various types of feedback  and coupling schemes.  In particular, we
use non-local feedback $s_{\delta}(x,t)$ and time-delayed feedback
$s_{\tau}(x,t)$. On the one hand, this is motivated by physiological
considerations (Fig.~\ref{fig:nvc}). On the other hand, these two
distinguished types allow for a clear separation of the effect of the
introduced space scale $\delta$ and time scale $\tau$ upon the
spatio-temporal pattern of SD. We have shown that these scales
$\delta$ and $\tau$ are closely related. The reason is, of cource,
that the feedback is applied to control of travelling pulses. The velocity
of SD relates space to time scales.  Their numerical values can be
compared after normalizing them with the pulse width $\Delta x$ and
pulse duration $\Delta t$, respectively. The common effect of both
types of feedback  on the traveling pulse is then reflected in the similar
location of optimal control around half the pulse width and pulse
duration for $s_{\delta}(x,t)$ and $s_{\tau}(x,t)$, respectively
(Fig.~\ref{fig:kDeltaTau}).  Therefore, the feedback mechanism, as
given in a general form in Eq.~(\ref{eq:s}), and illustrated in
Fig.~\ref{fig:signal}, provides a generic mechanism to modulate
excitability in reaction-diffusion systems even if we consider only
its two limit cases.

The reaction-diffusion system we have used to demonstrate the
modulation of excitability by feedback is the FHN model. It represents
excitable media exhibiting excitability of type~II. Furthermore, we
have constrained our investigations to the section $a=1$ and
$\gamma=0.5$ in the four-dimensional parameter space of this
model. Both restrictions, excitability of type II and this choice
within the parameter space, will now be discussed. Type~II
excitability is related to a Hopf bifurcation, which is characterized
by the fact that a periodic state bifurcates with a nonzero frequency
\cite{ERM98}.  Within an exponentially small range of either $\beta$
or $\epsilon$ after the Hopf bifurcation a transition occurs, called
{\em canard explosion}, from a state with small amplitude oscillations
to a large amplitude relaxation oscillation.  The canard explosion
produces the threshold behavior (all-or-none) of the system needed to
exhibit excitable behavior. Another typical bifurcation scenario
leading to excitability is that the periodic state appears at a zero
frequency \cite{WIL99a,ERM98}, called type I excitability. This can be
modeled by a saddle-node bifurcation on a limit cycle or saddle-node
infinite period bifurcation (SNIPER) \cite{HIZ06,HIZ07}.

It is not clear which type of excitability is the basis of
SD. However, experiments with retinal spreading depression indicate
excitability of type~II behavior \cite{DAH03b}. The extracellular
potassium concentration ($[K^+]_e$) was increased stepwise. Above a
usually well preserved ceiling level of $[K^+]_e=10\mbox{mM}$
\cite{HEI77} a rather spontaneous onset of oscillations with finite
period was observed. Due to the presence of noise and a rather large
step width of $\Delta[K^+]_e=2\mbox{mM}$, a definite conclusion is not
yet possible. However, we assume that our results hold in
principle also for excitable media in which each local element is of
excitability type~I. The difference is manifest only close to the
bifurcating state of the individual excitable element. In an excitable
media, that is in a spatially extended system, excitability is defined by the
distinguishing feature that the excited state breaks away from a local
stimulated area due to transport, which usually involves diffusion.  The
bifurcating state of the individual elements seems therefore less
important than the saddle-node bifurcation that leads to the emergence
of traveling pulses (Fig.~\ref{fig:pb}, inset {\sf b}).

The FHN system is the most generic type that shows excitability of type
II, because the activator equation, Eq.~(\ref{eq:udot}), has a cubic
nonlinearity, which is generic for bistability \cite{SCH72,SCH01}. Since we aim to
describe generic features, let us briefly comment on the choice of the
two parameters that we have fixed ($a=1$ and $\gamma=0.5$). With this
choice, the value of the diffusion coefficient in the main part of the
remaining section of the parameter space is between $5\cdot
10^{-6}\mbox{cm}^{2}\mbox{s}^{-1}$ and $100\cdot
10^{-6}\mbox{cm}^{2}\mbox{s}^{-1}$ (Fig.~\ref{fig:pb}). The diffusion
coefficient of $[K^+]_e$, a substance often related to the activator
or at least to the species by which SD propagates, is about $20\cdot
10^{-6}\mbox{cm}^{2}\mbox{s}^{-1}$ in aqueous solution \cite{STO50}.
There are two counteracting effects that can significantly change this
value in cortical tissue.  First, due to the porous geometrical
structure and small volume fraction of the exctracellular space, which
is similar to a soap phase, the apparent diffusion coefficient of
$[K^+]_e$ in brain tissue is estimated to $7.2\cdot
10^{-6}\mbox{cm}^{2}\mbox{s}^{-1}$ \cite{NIC81}.  Second, the
possibility that $[K^+]_e$ enters glial cells in one place and leaves
elsewhere provides a form of facilitated diffusion that is estimated
to be up to five times more important than diffusion in the
extracellular space \cite{GAR81}. In summary, a generic model that
simulates traveling pulses with a diffusion coefficient $\tilde{D}$ of
reasonable order of magnitude seems to be appropriate.

We propose that a failure of feedback provides a common mechanism of
the emergence of spreading depolarizations in migraine and stroke. A
recent study taking a complementary bottom-up approach, by describing
SD in a detailed biophysical model, has concluded that the key to
normal stability of cortical tissue is the effective regulation of
$[K+]_e$ by the neuron's Na-K ion pump and the glia-endothelial system
\cite{KAG00}.  This complements our conclusions.  In our generic
approach, our concern is not to identify the precise species behind
activator, inhibitor, and feedback signals, but rather to describe
spatial properties of their interaction and the emergence of
travelling pulses. On the contrary, a detailed biophysical model of
properties of a single neuron and its surrounding extracellular
compartments can only infer statements about the local excitable state.
Whether a local breakdown of ion and transmitter homeostasis recruits
further tissue into the excitable state or remains restricted to the
initial focus is the clinically important question. This can only be
answered in an extended system with sufficient spatial resolution,
where a pulse propagation boundary can be defined
(Fig.~\ref{fig:pb}). The existence of the propagation boundary has
consequences for therapeutic approaches that aim at an effective
reduction of excitability to hold the tissue below the bifurcating
state and with that limit the tissue at risk \cite{DAH07a}.

%Focal neocortical epilepsy is characterized by repeated,
%localized “interictal” events and rarer, full
%“ictal” events in which the activity breaks away
%from the restricted focus (Gastaut and Broughton, 1972). 

% The
%feedback along long range conditions could, for example, activate
%Na-K ion pump

\section*{\sf Acknowledgements}
This work was supported by DFG in the framework of Sfb 555. The
authors would like to thank H.\,Engel, H.\,R.\,Wilson,  and K.\,Showalter for fruitful discussions.

%\bibliographystyle{prwithtitle}
%\bibliography{ref}

%%%%%%%%%%%%%%%%%%%%%%%%%%%%%%%%%%%%%%%%%%%%%%%%%%%%%%%%%%%%%%%%%%%%
\end{document}